\documentclass[journal]{IEEEtran}

\ifCLASSINFOpdf
\else
   \usepackage[dvips]{graphicx}
\fi
\usepackage{url}
\usepackage{graphicx}
\usepackage{booktabs}
\usepackage{hyperref}
\usepackage{balance}

\begin{document}

\title{Self-Supervised Learning of \\Context-Aware Pitch Prosody Representations}

\author{Camille Noufi, Prateek Verma
\thanks{Submitted for review on 25 March 2021.}
\thanks{Camille Noufi is with the Center for Computer Research in Music and Acoustics (CCRMA), Stanford University, Stanford CA 94305 (email: cnoufi@ccrma.stanford.edu). Prateek Verma is with the Stanford AI Lab, Stanford University, Stanford CA 94305 (email: prateekv@stanford.edu).}}

\markboth{IEEE Signal Processing Letters, Vol. X, No. X, Month 20XX}
{Shell \MakeLowercase{\textit{et al.}}: Bare Demo of IEEEtran.cls for IEEE Journals}

\maketitle

\begin{abstract}
In music and speech, meaning is derived at multiple levels of context. Affect, for example, can be inferred both by a short sound token and by sonic patterns  over a longer temporal window such as an entire recording. In this letter, we focus on inferring meaning from this dichotomy of contexts. We show how contextual representations of short sung vocal lines can be implicitly learned from fundamental frequency ($F_0$) and thus be used as a meaningful feature space for downstream Music Information Retrieval (MIR) tasks. We propose three self-supervised deep learning paradigms which leverage pseudotask learning of these two levels of context to produce latent representation spaces. We evaluate the usefulness of these representations by embedding unseen pitch contours into each space and conducting downstream classification tasks. Our results show that contextual representation can enhance downstream classification by as much as 15\% as compared to using traditional statistical contour features.
\end{abstract}


\IEEEpeerreviewmaketitle

%%%%% INTRO %%%%%%%%%%%%%%%%%%%%%%%%
\section{Introduction}
\label{sec:introduction}

Growing interest within the music information retrieval (MIR) community in utilizing ``in the wild" datasets is shedding light on the challenge of securing the high quality, expert labeling needed for supervised deep learning models to extract contextual and complex musical meaning.

In this letter, we learn local and broad contextual representation spaces that influence a singer's prosodic pitch pattern (measured by the fundamental frequency $F_0$ extracted from a monophonic vocal signal). The $F_0$ contour -- the \textit{continuous} trajectory of $F_0$ in time, also often referred to as `pitch contour' -- is a commonly used feature for numerous MIR problems including instrument recognition in monophonic or polyphonic recordings \cite{bittner2017-pcMidlevel}, melody extraction in polyphonic music \cite{salamon2012statistical, Salamon2012-PC}, genre and style labeling \cite{panteli2017-worldMusic,Muller2019-F0Fusion}, classification of affect in the singing voice \cite{emotion-acoustic-patterns} and study of intonation patterns in the singing voice \cite{Dai2019IntonationTW}. The success of this feature in enabling such MIR tasks has prompted the creation of robust and accurate $F_0$ extraction methods such as CREPE \cite{kim2018crepe}, as well as the development of $F_0$ contour classification libraries Melodia \cite{Salamon2012-PC}, BITELLI \cite{bittner2017-pcMidlevel,panteli2017-worldMusic} and PyMus \cite{Muller2019-F0Fusion}. These tools allow for the fundamental frequency $F_0$ to be disentangled from the rest of the audio spectrum.

For the singing voice, one can hypothesize that the structure of an $F_0$ contour is directly influenced by 1) the preceding and following contours in a vocal line and 2) the structural characteristics present across an entire audio recording. These spans of temporal influence are defined in this paper as local (contiguous) context and broad (file-level) context, respectively. Furthermore, the aforementioned applications that leverage $F_0$ contours typically utilize statistical summaries of contour structure as a mid-level feature as input for relevant classification tasks. This mapping suggests that context is implicitly encoded into the structure of the $F_0$ contour, and thus that there exists an inverse mapping between the immediate and broad temporal contexts of a vocalization and its $F_0$ contour structure. Thus, $F_0$ can be leveraged as a data source for MIR tasks when uncontrolled variability in recording (e.g microphone, recording environment, unknown preprocessing, etc) confound analysis. In such datasets where the full spectrum cannot be reliably utilized, it would be helpful to maintain access to broader information encoded within the $F_0$ contour. However, little prior investigation has been conducted on the influence of context on vocal pitch prosody. 

We aim to learn these contexts in a self-supervised manner and then apply the learned-context space as a feature for downstream classification tasks.  Self-supervised learning leverages labels that are naturally part of the training data, rather than requiring separate external labels. Using these implicit labels, a self-supervised model learns information on the structure of the input. This training paradigm is thus optimally suited for learning how low-level features relate to one another. In computer vision, various self-supervised learning tasks (e.g., colorization, relative position) have been proposed for learning visual representations \cite{doersch2017multi}. Additionally, self-supervised learning has been leveraged in natural language processing (NLP) to learn semantic meaning in text. The Word2vec \cite{Mikolov2013-w2v} embedding scheme has been particularly popular, having influenced methods for speech recognition \cite{Chung2018-Speech2Vec} and, in music, learning functional tonal harmony \cite{chord2vec,music2vec}. These algorithms use surrounding context (e.g., adjacency of words in a sentence or document) rather than structure (e.g., spelling) to learn the meaning of data. Chung and Glass show that word embeddings functioned more semantically with the incorporation of audio features into an experiment similar to that in Word2Vec \cite{Chung2018-Speech2Vec}. Although their results were oriented toward boosting performance of word embeddings, the results support previous research suggesting that vocal intonation provides additional semantic meaning \cite{speechmusiccode}. 

The contributions of this letter are as follows:

i) We learn latent representations of pitch contour embeddings from $F0$ data without access to external labels and use these representations to solve several downstream tasks. We find competitive results on a different dataset, indicating that this method can generalize.\newline
ii) We compare self-supervised latent representations of different contexts, and show how combining contextual features with statistical features often better solves a downstream task than using a single self-supervised embedding or a statistical summary as a feature.\newline
iii) We show how our latent representations discerns high-level characteristics relating to pitch prosody, and gives poor results on characteristics that do not (e.g, vowels), providing initial insight into the applicability of these representations. 

\section{Self-Supervised Model Design}\label{sec:methods}
We design three self-supervised learning methods to create three different representation spaces. For each method, a pseudotask is used to learn the relationship between the pairs of $F_0$ contours embedded into these spaces. Following the creation of these representations spaces, unseen $F_0$ contours are passed through these encoders, and their embeddings into these spaces are then used as the data source for a variety of downstream subtasks. Similar frameworks have been widely used in computer vision, where one learns suitable representation spaces, followed by a  simple classifier for  downstream classification subtasks \cite{chen2020simple}. Solving the pseudotask is a separate system from the subtask in which we are interested. The applicability of the learned representation spaces to the separate subtasks of interest is discussed in Section \ref{sec:results}.

\begin{figure}[!hbt] \centering \includegraphics[width=0.95\linewidth]{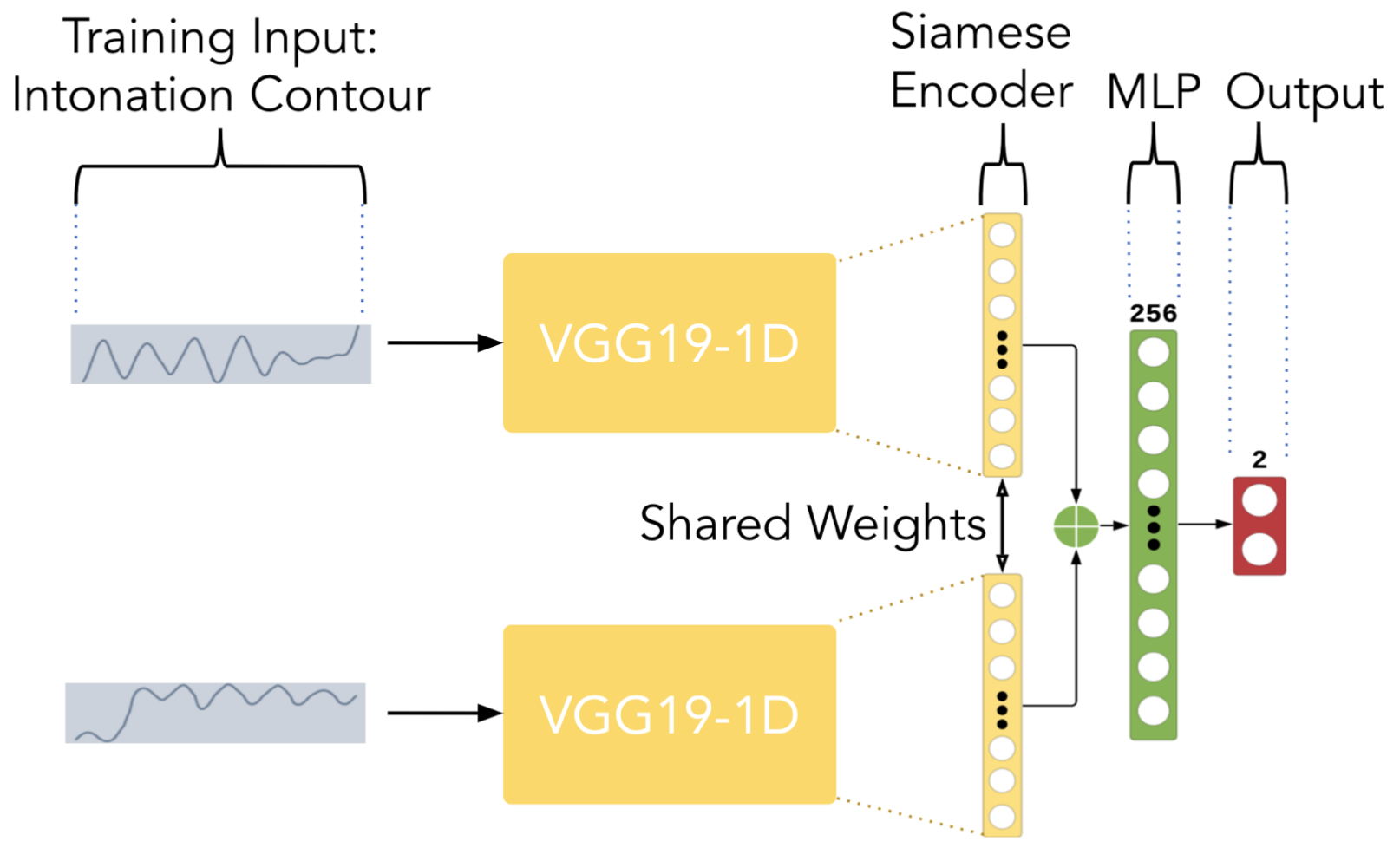} \caption{Siamese VGG19-1D neural network architecture used for binary pseudotask learning.} \label{fig:snn} \end{figure}  

\begin{figure}[!hbt] \centering \includegraphics[width=0.95\linewidth]{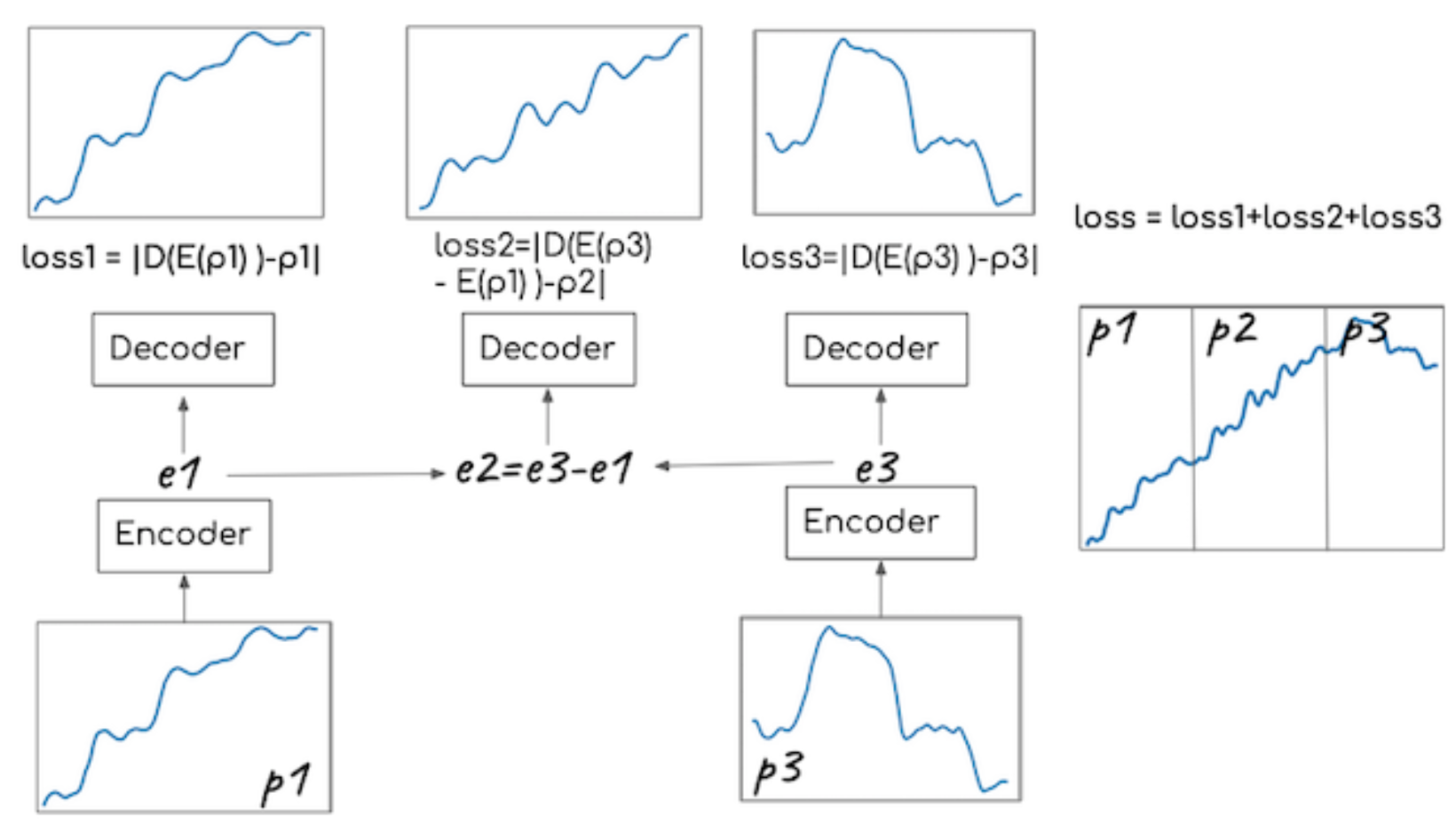} \caption{ Slot-filling MLP encoder(\(E\))-decoder(\(D\)). \(e_1, e_2, e_3\) are the encoding spaces of three contiguous $F_0$ contours \(p_1, p_2, p_3\).} \label{fig:slot-fill}\end{figure}

\begin{figure*}[!hbtp] \centering \includegraphics[width=\textwidth]{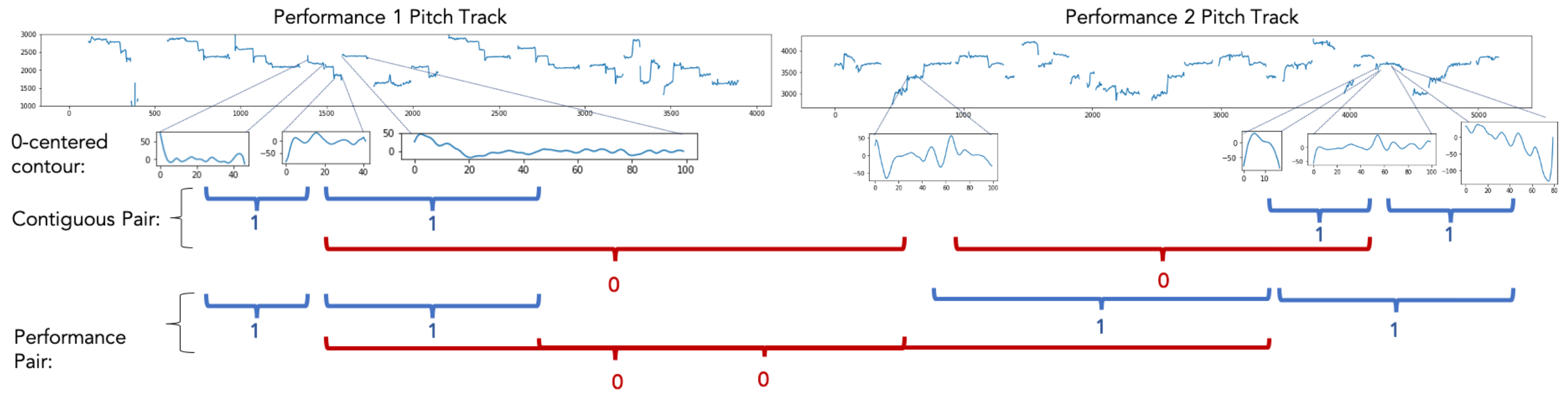} \caption{Contour selection for the Contiguous and File-level context models.  In the contiguous pairing scheme, adjacent contours are extracted and labeled $Y=1$ (blue) while contours separated in time are labeled $Y=0$ (red).  In the file-level pairing scheme, contours from the same performance/recording are labeled $Y=1$ while contours from different recordings are labeled $Y=0$.} \label{fig:contour-pair} \end{figure*}

\subsection{Binary Pairing Criterion}\label{criterion} The binary criterion labels are derived implicitly from the training data. If a pair of pitch contours meets a specified criterion, they are given the label $Y=1$, otherwise they are given the label $Y=0$. Our first pseudotask aims to learn whether a random pair of pitch contours exist within the same recording or not (which we label respectively, $Y=1$ and $Y=0$). Our second pseudotask is to learn whether two $F_0$ contours are contiguous. Contours directly adjacent to each other are considered a contiguous pair and are assigned the label $Y=1$. For non-contiguous contours, we select half of the negative pairs (we which label as $Y=0$) to be from the same recording, with any temporal separation between the pair allowed. For the other half of negative pairs, we ensure the pairs originate from different files. Figure \ref{fig:contour-pair} provides a visual example of the binary pairing schemes.

\subsection{Binary Pairing Model Architecture}
We propose learning a contextual representation by training a Siamese neural network (SNN) to determine whether or not a pair of $F_0$ contours meet the pairing criterion. As input to the SNN, we use random pairs of $F_0$ contours created using the method described below in Section \ref{datasets}. The SNN consists of twin 1D variants of the VGG architecture \cite{Vggish} (denoted here as VGG19-1D). We use the 128-dimensional dense layer following the VGG19-1D convolutional layers as our representation space. We join the VGG19-1D twin networks by concatenating the twin dense layers, and passing them to another fully connected 256-dimension layer. The final layer performs binary classification using soft-max activation and cross-entropy loss between the output of the model and the pseudo-label of the contour pair as described in Section \ref{criterion}. Figure \ref{fig:snn} visualizes the architecture of this SNN. We choose to work with the VGGish architecture (rather than, e.g., ResNet-50) due the reduced model complexity, thus providing faster training. A comparison of feasible encoder architectures is beyond the scope of the current work.

\subsection {Slot-Filling Design} Our third pseudotask is to learn a sequence-modeling task. Given two contour segments, we learn a representation space that is used to predict an unknown contour directly between the two. We make an assumption that the representative space of a third $F_0$ contour is a vector addition of the encodings of the previous two. We learn latent codes \(e_1, e_2, e_3\) for three consecutive pitch contours \(p_1, p_2, p_3\). We train our encoder \(E\) and decoder \(D\) with shared weights to minimize the reconstruction loss for \(e_1\) and \(e_3\). \(D\) reconstructs \(p_2\) from a vector subtraction of the encodings i.e. $p_2^{'} = D(e_3-e_1)$, where each of the encoded latent vectors \(e_i\) = \(D\)\((p_i)\). Via this design, most of the heavy lifting is done by the encoder-decoder to map $F_0$ contours to a space where a simple operation such as vector addition is sufficient to decode what is next. The encoder and decoder are both 3-layer multilayer perceptron networks (MLP), where the first two layers of each use 2048 neurons, the bottleneck of the encoder is size 20, and the output of the decoder is size 100. Such MLP-based architectures have been successful in encoding salient characteristics of short audio signals \cite{verma2016frequency}\footnote{This method differs from contrastive predictive coding \cite{oord2018representation}. We learn representations that can decode from the latent space the original contour while using the same latent code to predict the next via addition, rather than predicting using a recurrent network.}. Figure \ref{fig:slot-fill} visualizes this architecture. 

\section{Datasets and Implementation} \label{datasets}

VocalSet \cite{vocalset}, a dataset comprised of recordings of 11 male and 9 female singers, is used to train the models learning each pseudotask. The set includes ten different vocal techniques in contexts of scales, arpeggios and phrases. The dataset is split into 2,684 training tracks sung by 7 females and 8 males, and 872 test pitch tracks sung by 2 females and 3 males. In order to judge the generalization of the representation spaces on downstream classification tasks, the Song portion of The Ryerson Audio-Visual Database of Emotional Speech and Song (RAVDESS) \cite{ravdess} is used alongside the test portion of VocalSet. This portion includes 1,012 short audio recordings sung by 12 males and 12 females. Both datasets provide metadata for each recording. VocalSet provides the Gender, Singer, Phrase Type being sung in the recording, Vocal Technique used, and Vowel. RAVDESS provides Gender, Actor, Emotion, Emotional Intensity (normal or strong), and Statement ("kids are talking by the door" or "dogs are sitting by the door").
Pitch tracks -- $F_0$ annotations over the entire audio file -- are extracted via CREPE \cite{kim2018crepe} using a frame size of 12 ms. Each pitch track is segmented into smaller contours. A contour length of 100 frames (120 ms) is selected to be the maximum duration and the input size into our downstream models. Contours greater than 100 frames are split and contours less than 100 frames are zero-padded. Frequencies are converted to cents and each contour is transposed to have a median value of 0 cents to reduce salience of a singer's vocal range. Random shifting by a maximum of 1,200 cents is applied as a form of data augmentation. 

All self-supervised models were trained between 30-100 epochs using a batch size of 50 contour pairs. Adam optimization was used with a learning rate tuned from 1e-4 to 1e-7 in steps of 0.1 using a patience threshold of 5 epochs. Adam optimization was used Training was carried out using Tensorflow on P100/V100 GPUs within the Google Cloud computing environment.

% Please add the following required packages to your document preamble:
% \usepackage{booktabs}
% \usepackage{graphicx}
\begin{table*}[]
\centering
\caption{Ablation study results of three pseudotasks: Validation accuracy of File-Pair-Context and Contiguous-Pair-Context models, and $\ell2$-loss of the Slot-Filling model, column 1. Test-set accuracy (\%) of downstream classification subtasks performed with a given embedding. Subtask results are shown for the PyMus statistical features, the context-based embeddings and a combination of the PyMus features with the embeddings (ex. File-Contig.-PyMus).}
\label{tab:class-results}
\resizebox{\textwidth}{!}{%
\begin{tabular}{@{}lccccccccccc@{}}
\toprule
 &  &  &  & VocalSet & \multicolumn{1}{l}{\cite{vocalset}} &  &  &  & RAVDESS & \multicolumn{1}{l}{\cite{ravdess}} &  \\ \midrule
Embedding/Feature & \multicolumn{1}{c|}{PT Perf.} & \multicolumn{1}{c|}{Gender} & Singer & Phrase & Technique & \multicolumn{1}{c|}{Vowel} & Gender & Actor & Emotion & Statement & Intensity \\ \midrule
PyMus & \multicolumn{1}{c|}{-} & \multicolumn{1}{c|}{65.0} & 38.3 & 66.6 & 38.9 & \multicolumn{1}{c|}{19.4} & 59.5 & 23.8 & 22.2 & 52.2 & 55.2 \\ \midrule
File-Context & \multicolumn{1}{c|}{81\%} & \multicolumn{1}{c|}{74.5} & 52.9 & 79.4 & 41.9 & \multicolumn{1}{c|}{21.1} & 63.9 & \textbf{38.4} & \textbf{35.4} & 54.2 & \textbf{57.7} \\
Contig.-Context & \multicolumn{1}{c|}{99\%} & \multicolumn{1}{c|}{64.5} & 42.2 & 72.9 & 30.6 & \multicolumn{1}{c|}{19.2} & 58.3 & 20.2 & 19.1 & 50.2 & 55.5 \\
Slot-Filling & \multicolumn{1}{c|}{$\ell2=0.03$} & \multicolumn{1}{c|}{58.1} & 40.6 & 52.3 & 22.1 & \multicolumn{1}{c|}{16.4} & 51.3 & 30.9 & 23.3 & 49.9 & 51.4 \\
File-PyMus & \multicolumn{1}{c|}{-} & \multicolumn{1}{c|}{77.0} & 54.5 & \textbf{81.1} & \textbf{45.1} & \multicolumn{1}{c|}{20.3} & 65.1 & 22.5 & 26.6 & 54.2 & 57.6 \\
Contig.-PyMus & \multicolumn{1}{c|}{-} & \multicolumn{1}{c|}{74.2} & 45.1 & 76.7 & 39.9 & \multicolumn{1}{c|}{19.9} & 62.4 & 26.2 & 22.2 & 52.2 & 58.0 \\
File-Contig. & \multicolumn{1}{c|}{-} & \multicolumn{1}{c|}{67.2} & 44.5 & 77.3 & 39.6 & \multicolumn{1}{c|}{25.1} & 64.5 & 31.4 & 22.2 & \textbf{56.6} & 55.8 \\
File-Contig.-PyMus & \multicolumn{1}{c|}{-} & \multicolumn{1}{c|}{\textbf{77.2}} & \textbf{58.2} & 79.6 & 43.0 & \multicolumn{1}{c|}{24.5} & \textbf{70.3} & 22.1 & 19.3 & 56.5 & 58.6 \\
Chance & - & 50.0 & 20.0 & 16.7 & 5.9 & 16.7 & 50.0 & 20.0 & 16.7 & 50.0 & 50.0 \\ \bottomrule
\end{tabular}%
}
\end{table*}

%%%%%%%%%%%%%%%%%%%%%%%%%%%%%%% Results %%%%%%%%%%%%%%%%%%
\section{Evaluation}\label{sec:evaluation}

The training results of the three models are shown in Table \ref{tab:class-results}. The SNN identifies pairs from the same file with 81\% accuracy and contiguous pairs with 99\% accuracy. The slot filling model is optimized over a validation set and able to recreate the center contour $e_2$ with an $\ell2$-loss of 0.03. The latent embedding spaces of these three trained models are the focus of evaluation.

We extract 10,000 random 100-frame contours from the VocalSet test partition and the RAVDESS Song dataset and embed these contours into the three latent representation spaces. These embeddings are used as the input feature vectors to the simple classification models. As our baseline, we use the PyMus contour feature extraction library, introduced in \cite{Muller2019-F0Fusion}, to extract 17 statistical summary features of each contour. These features measure modulation, fluctuation, and the average gradient of each $F_0$ contour using both cents and fundamental frequency. We examine the effect of combining the embeddings of different contexts together, as well as combining the context-based embeddings with the hand-crafted statistical features. As far as we know, there doe not yet exist other self- or unsupervised feature learning methods against which we can compare our proposed design. We follow the conventions of the recent related works in the field of self-supervised representation learning (\cite{chen2020simple} and \cite{nagrani2020disentangled}) by reporting downstream task classification accuracy as our performance metric. This also compares utility of different learned representation for various tasks that had not been explored before in the literature. 

We train a 3-layer multilayer perceptron (MLP) network on the embeddings and hand-crafted features. The network consists of 128 nodes per layer. The first two layers are followed by a ReLU activation. The archtecture of the MLP is consistent across all tasks. We normalize the input features and undersample in cases  of class imbalance. Training uses Adam optimization and an adaptive learning rate beginning at 1e-3. Five-fold cross validation is used to obtain averaged accuracy and F1 scores for each category. Hyperparameter and architecture tuning of the subtask is not explored in this work.

\section{Results and Discussion}\label{sec:results}

Subtask accuracies using the representation spaces generated by the pseudotasks, shown in Table \ref{tab:class-results}, are equal to or greater than those learned from the hand-crafted features. We find that the representations' usefulness as a feature space differs depends on the goal. Notable boosts in performance are seen for Singer, Actor, Phrase Type, Vocal Technique, and Emotion, with up to 15\% increase in performance. As a sanity check, we expect to find Vowel accuracy and Intensity to be on par with chance, as Vowel is defined by weight of upper harmonics in the spectrum rather than by $F_0$. Similarly, difference in Statement within RAVDESS relies mostly on phonetic variation and Intensity is largely a product of energy amplitude and distribution within the frequency spectrum. Conversely, the file-level embeddings are able to predict gender and actor/singer at least as well as the baseline features, suggesting that short patterns of intonation existing repeatedly across a recording can be recognized and related to learned pronunciation style and prosodic tendencies \cite{Dai2019IntonationTW}. Additionally, this result suggests that context-aware embeddings can provide information about high-level attributes typically linked to spectral features, like Gender and Emotion, from context of $F_0$. 

The best performance often results when contextual embeddings and the hand-crafted features are combined. The embedding space created via the file-level pairing pseudotask enables the highest classification accuracy for the subtasks. Although pseudotask accuracy on learning contiguous pairs was 99\%, that accuracy did not translate to the resulting representation being the most useful input feature. Despite the slot filling model's success at reproducing contours of similar structure, the resulting representation space did not serve as a more useful feature than the other embeddings or the baseline. This result indicates that the slot-filling representation space failed to capture significant contextual information not available in the contours' structural content. Thus, combination of the slot-filling representation with the other representations and hand-crafted features was not further explored within this experiment.  This suggests that learning a pseudotask successfully, as judged by a given accuracy metric, does not necessarily correlate with the performance of the resulting representation space in downstream tasks. A current hypothesize as to why the file-level pseudotask yielded the most useful space is akin to why the Word2vec variants mentioned in Section \ref{sec:introduction} are successful. Multiple instances of pairs from the same file are used in training, making the ratio of ``corpus size" to ``vocabulary" size much larger than in the contiguous pair training scheme, where exact replicas of contiguous pairs are rare.

\section{Summary and Future Work}

As evidenced by the mismatch between pseudotask accuracy and downstream task performance, a focus on evaluation of the interpretability of latent space is necessary. Additionally, a multi-task loss combining the proposed methods can be explored.  It also will be worthwhile to explore transformer-based encoders in future work. 

In summary, we proposed three self-supervised deep neural network architectures to learn contextual information relating to sung vocal contours directly from their relationship to surrounding contours without external annotation. These training tasks produce representation spaces that can be used as features for downstream MIR tasks that would benefit from contextual information but lack metadata or other relevant labels. Initial evaluations of these embeddings demonstrate that they enhance the performance of downstream classification by several points as compared to learning on hand-crafted contour features alone. The results indicate that different contextual spaces are suited for different subtasks. The usefulness of file-level context, alone or in combination with local context and structural features, suggests that a combination of broad and local information provides the best representation of a sung vocal line.  

\section*{Acknowledgments}
The authors thank Dr. Jonathan Berger for his insights and feedback on the writing of this letter.

\balance
\bibliographystyle{IEEEbib}
\bibliography{bib}

\end{document}